\documentclass[twocolumn,showpacs,prx,superscriptaddress,amsmath,amssymb]{revtex4}

\usepackage{booktabs}
\usepackage{color}
\usepackage{graphicx}
\usepackage{dcolumn}
\usepackage{bm}
\usepackage{graphicx}

\begin{document}

\title{Compton Scattering Energy Spectrum for Si and Ge Systems}

\author{Chen-Kai Qiao}
\email{chenkaiqiao@126.com}
\affiliation{College of Physics, Sichuan University, Chengdu, Sichuan, 610065}

\author{Hsin-Chang Chi}
\affiliation{Department of Physics, National Dong Hwa University, Shoufeng, Hualien, Taiwan 97401}

\author{Shin-Ted Lin}
\affiliation{College of Physics, Sichuan University, Chengdu, Sichuan, 610065}

\author{Peng Gu}
\affiliation{College of Physics, Sichuan University, Chengdu, Sichuan, 610065}

\author{Shu-Kui Liu}
\affiliation{College of Physics, Sichuan University, Chengdu, Sichuan, 610065}

\author{Chang-Jian Tang}
\affiliation{College of Physics, Sichuan University, Chengdu, Sichuan, 610065}

\date{\today}

\begin{abstract}
In the present work, we study the atomic Compton Scattering which could have great impacts on dark matter direct detection experiments. We give a quantitative analysis of the Compton scattering energy spectrum for Si and Ge atomic systems. The theoretical results on Compton scattering are calculated within the frameworks of free electron approximation (FEA) and relativistic impulse approximation (RIA). The low-energy transfer and near photoionization threshold regions are especially considered in this work. In RIA calculations, to obtain the atomic ground states, we adopt an \emph{ab initio} calculation in the fully relativistic Dirac-Fock theory.
\end{abstract}


\maketitle


\section{Introduction \label{sec:1}}

Atomic Compton Scatterings have been extensively investigated and widely applied in many areas of physics over the past several decades \cite{Cooper1,Cooper2,Cooper3,Ji,Porter,Wang,Phuoc}, and the development of modern gamma-ray spectrometer and imaging devices are also benefit a lot from the Compton Scatterings \cite{Takada,Mihailescu,Chiu}.

Recently, dark matter problem is one of the most attractive topics among interdisciplinary branches of science. The preponderant evidences for existence of dark matter have been found from the special rotation curves of spiral galaxies \cite{Zwicky,Rubin,Corbelli}, gravitational lensing \cite{Clowe}, large scale structure formation of the universe \cite{Blumenthal,Davis}, cosmic microwave background and baryon acoustic oscillations \cite{WMAP,Planck,Undagoitia}. There are three primary approaches that are direct detection, indirect detection and accelerator detection to search these unknown dark matter particles. These detection experiments strictly constrain the mass of dark matter particles, strength of interactions, as well as other parameters over past few years \cite{Undagoitia,Bertone,RDG2016,RDG2018}.

Compton scattering plays a key role in direct detection of dark matter particles, for it is one of the nonnegligible background processes for X-ray and gamma-ray radiations that must be eliminated in experiments \cite{Barker,Ramanathan,Robinson}. In dark matter detection experiments, the behavior of the detector responses of interactions between dark matter particles $\chi$ and target atoms are quite similar with those of low-energy transfer atomic Compton scatterings. On the one hand, many underground experiments are focused on the elastic $\chi-N$ scattering, aiming to detect these unknown particles through nuclear recoil signals. However, in experiments like CDEX and CDMSlite, the nuclear recoils produced by $\chi-N$ scattering can hardly be distinguished from the electron recoils produced by atomic Compton scattering, especially in regions near the photoionization thresholds. On the other hand, dark matter particles can also cause electron recoils in the detectors. In particular, they can directly interact with atomic electrons through the $\chi-e$ scattering \cite{Bernabei,Agnes}, or they can cause electron ionizations during scattering with atomic nuclei \cite{Moustakidis,Vergados,Ejiri,Ibe}. The energy deposition spectra of these processes are very similar to those of atomic Compton scatterings, which makes Compton scattering to be nonnegligible backgrounds. Furthermore, the Compton scattering processes can affect the low-energy calibration of these experiments. Therefore, the study of dynamical properties and the detector response of Compton scattering may give us guidance on the background understanding, background analysis, background subtraction and suppression, as well as the identification of dark matter particles.

Compton scattering is conventionally approached from the free electron approximation (FEA) and Klein-Nishina formula \cite{Klein-Nishina,Sakurai}. In FEA, the electron interactions with atomic ions are neglected, and electrons are also assumed to be at rest prior to photon scatterings in the laboratory frame. The FEA and Klein-Nishina formula works well in the high-energy transfer region, where atomic binding effects are negligible and the atomic electrons are asymptotically free. However, in the low-energy transfer region, in which cases would have great impacts on the dark matter detections \cite{Ramanathan}, the atomic binding effects become dominant, the FEA breaks down and more advanced approaches are needed. The atomic binding effects and the pre-collision motion of electrons can be treated effectively in the relativistic impulse approximation (RIA) developed in previous years \cite{Eisenberger1,Eisenberger2,Ribberfors1,Ribberfors2,Ribberfors3,Ribberfors4}. With the advantages of simplicity and flexibility, the RIA formulation has been widely applied to atomic \cite{Ribberfors3}, condensed matter \cite{Kubo,Rathor,Pisani,Aguiar}, nuclear \cite{Brusa}, and elementary particle physics \cite{Ramanathan}. In particular, in the conventional Monte Carlo simulation program Geant4 \cite{Geant4}, which are extensively used in nuclear and elementary particle physics, the Compton scattering processes are treated using the ideas of FEA and RIA \cite{Geant4b,Livermore,Monash}.

Although the ideas of RIA are adopted into various simulation models of the current Geant4 program, the kinematics of Compton scattering is approximately treated in some simulation models \cite{Geant4b,Livermore}. Further, the atomic databases relevant to the Compton scattering, which were calculated by Biggs \emph{et al.} and Hubbell \emph{et al.} in many years ago \cite{Biggs,Hubbell,EPDL}, are outdated. These old databases, which are calculated using the nonrelativistic Hartree-Fock (HF) theory, are insufficient to give a precise prediction and to meet the current requirements. Therefore, an \emph{ab initio} calculation of atomic Compton scattering process considering the relativistic, atomic binding, and other many-body effects is needed.

In this study, we carry out a concrete analysis on the Compton scattering energy spectrum in the FEA and RIA formulations. The low-energy transfer and near photoionization threshold regions are especially considered in this work. We mainly focus ourselves on the Compton scattering of the Si and Ge atomic systems, irrespective of the influences coming from complex geometrical structure of detectors. With sufficiently low threshold and high efficiency, Si and Ge are ideal materials for experimental detection of charged and neutral particles, especially for dark matter direct detection experiments \cite{CoGeNT,CDMS,CDEX,CDMS2}. In dark matter direct detections, there is an irreducibly undesirable background called as the ``neutrino floor'' \cite{Monroe,Billard}. In order to take the atomic binding and relativistic effects into consideration, we utilize the fully relativistic Dirac-Fock formalism to calculate the ground states of Si and Ge atoms \cite{Grant0,Desclaux,Grant,Visscher}. In the low-energy transfer or near photoionization threshold regions, the Compton scattering energy spectrum presents several linear platforms and step structures approximately. Therefore, we adopt a linear fit approach of the Compton scattering energy spectrum in the low-energy transfer region for various incident photon energies. The slopes for K and L platforms and the relative altitude ratio between platforms on two sides of K and L steps at their photoionization thresholds are carefully studied in this work, and we compare our theoretical calculations with the Monte Carlo simulations \cite{Gu}. An experiment aimed to text our results is ongoing \cite{Experiment}.

This paper is organised as follows: Section \ref{sec:2} gives a description of the theoretical framework used in this work. Section \ref{sec:3} is devoted to results and discussions. Summary and conclusions are given in Section \ref{sec:4}.

\section{Theoretical Framework \label{sec:2}}

In this section, we give a brief introduction of the theoretical framework used in the present work. The energy spectrum of Compton scattering is calculated in FEA and RIA formulations, respectively.

\subsection{Free Electron Approximation \label{sec:2a}}

In the free electron approximation (FEA) and the Klein-Nishina formula, the electrons in Compton scattering are treated as free electrons, irrespective of the atomic environments in detectors. The atomic binding effects and electron correlations are neglected in the scattering process. Further, we assume that the electrons are at rest before the Compton scattering. In this formulation, the energy of the scattered photon $\omega_{f}$ is totally determined by its scattering angle $\theta$ via
\begin{equation}
\omega_{f}=\omega_{C}=\frac{\omega_{i}}{1+\omega_{i}(1-\cos\theta)/m_{e}c^{2}}
\end{equation}
Here, $m_{e}$ is the mass of electron, and $\omega_{C}$ is called as the Compton energy in literatures. When the scattering angle gets $\theta=180^{\text{o}}$, the energy of scattered photon reaches its minimum, meanwhile the energy transfer $T=\omega_{i}-\omega_{C}$ gets its maximum. In this case,
\begin{eqnarray}
\omega_{C}^{\text{min}} & = & \omega_{f}^{\text{min}}=\frac{\omega_{i}}{1+2\omega_{i}/m_{e}c^{2}}
\\
T^{\text{max}} & = & \omega_{i}-\omega_{C}^{\text{min}}=\omega_{i}-\frac{\omega_{i}}{1+2\omega_{i}/m_{e}c^{2}}
\end{eqnarray}
and it corresponds to the Compton edge in the Compton scattering energy spectrum.

In FEA formulation, the differential cross section of Compton scattering is given by \cite{Sakurai}
\begin{equation}
\bigg(
  \frac{d\sigma}{d\Omega_{f}}
\bigg)_{\text{FEA}}
=
\frac{r_{0}^{2}}{2}
\bigg(
  \frac{\omega_{C}}{\omega_{i}}
\bigg)^{2}
\bigg(
  \frac{\omega_{i}}{\omega_{C}}+\frac{\omega_{C}}{\omega_{i}}-\sin^{2}\theta
\bigg)
\end{equation}
and the scattering spectrum can be expressed as
\begin{eqnarray}
\bigg(
  \frac{d\sigma}{dT}
\bigg)_{\text{FEA}}
& = &
\pi r_{0}^{2} \frac{m_{e}c^{2}}{\omega_{i}^{2}}
\bigg[
  \frac{\omega_{i}}{\omega_{C}}
  +\frac{\omega_{C}}{\omega_{i}}
  -2m_{e}c^{2}
    \bigg( \frac{1}{\omega_{C}}-\frac{1}{\omega_{i}} \bigg) \nonumber
\\
&   &
  +m_{e}^{2}c^{4}
    \bigg( \frac{1}{\omega_{C}}-\frac{1}{\omega_{i}} \bigg)^{2}
\bigg]
\end{eqnarray}
where $r_{0}$ is the classical radius of electron, and $T=\omega_{i}-\omega_{C}$ is the energy transfer of the Compton scattering.

The FEA formulation works well only in cases when the atomic binding energies are negligible and the atomic electrons are approximately free, namely when the incident photon energy $\omega_{i}$ and energy transfer $T$ are sufficiently high. In the low-energy transfer region or the near photoionization threshold region, the FEA formulation becomes invalid and fails to fit the experiments \cite{Pratt1}.

\subsection{Relativistic Impulse Approximation \label{sec:2b}}

In the relativistic impulse approximation (RIA), the atomic binding effects and electron correlations are effectively considered, and the electron pre-collision motions prior to scattering are also included. It is these manybody effects that makes the RIA results significantly different from the FEA results. In RIA formulation, the energy of scattered photon can not be totally determined by its scattering angle as in FEA, and the Compton edge energy $\omega_{C}^{\text{min}}$ in Compton scattering energy spectrum no longer exist.

In this formulation, the differential cross section of Compton scattering is given by \cite{Ribberfors3,Ribberfors4,Brusa}
\begin{equation}
\bigg(
  \frac{d^{2}\sigma}{d\omega_{f}d\Omega_{f}}
\bigg)_{\text{RIA}}
=
\frac{r_{0}^{2}}{2}\frac{m_{e}}{q}
\frac{m_{e}c^{2}}{E(p_{z})}
\frac{\omega_{f}}{\omega_{i}}
\overline{X}(p_{z})
J(p_{z})
\label{RIA}
\end{equation}
and the function $\overline{X}(p_{z})$ is defined as
\begin{eqnarray}
\overline{X}(p_{z})
& = &
\frac{K_{i}(p_{z})}{K_{f}(p_{z})}
+\frac{K_{f}(p_{z})}{K_{i}(p_{z})} \nonumber
\\
&   &
+2m_{e}^{2}c^{2}
  \bigg(
    \frac{1}{K_{i}(p_{z})}-\frac{1}{K_{f}(p_{z})}
  \bigg) \nonumber
\\
&   &
+m_{e}^{4}c^{4}
  \bigg(
    \frac{1}{K_{i}(p_{z})}-\frac{1}{K_{f}(p_{z})}
  \bigg)^{2}
\label{function X Pmin}
\end{eqnarray}
with
\begin{eqnarray}
K_{i}(p_{z}) & = & \frac{\omega_{i}E(p_{z})}{c^{2}}+\frac{\omega_{i}(\omega_{i}-\omega_{f}\cos\theta)p_{z}}{c^{2}q}
\\
K_{f}(p_{z}) & = & K_{i}(p_{z})-\frac{\omega_{i}\omega_{f}(1-\cos\theta)}{c^{2}}
\\
E(p_{z}) & = & \sqrt{m_{e}^{2}c^{4}+p_{z}^{2}c^{2}}
\end{eqnarray}
In the above expressions, $q$ is the modulus of the momentum transfer vector $\boldsymbol{q} \equiv \boldsymbol{k}_{f}-\boldsymbol{k}_{i}$ and $p_{z}$ is the projection of the electron's initial momentum on the momentum transfer direction
\begin{equation}
p_{z} = \frac{\boldsymbol{p}\cdot\boldsymbol{q}}{q}
      = \frac{\omega_{i}\omega_{f}(1-\cos\theta)-E(p_{z})(\omega_{i}-\omega_{f})}{c^{2}q} \label{projection momentum}
\end{equation}

In the above expressions, the correction factor $J(p_{z})$ in differential cross section is called as the atomic Compton profile \cite{Biggs}
\begin{equation}
J(p_{z})\equiv \int\rho(\boldsymbol{p})dp_{x}dp_{y} \label{Compton profile}
\end{equation}
with $\rho(\boldsymbol{p})$ to be the ground state electron momentum density of the atomic systems. The Compton profile reflects the electronic properties of the atomic or molecular systems, and it has been widely studied in atomic and condensed matter physics \cite{Kubo,Rathor,Pisani,Aguiar}. For most of the atomic systems, the momentum distribution is spherical symmetric, then the atomic Compton profile reduces to
\begin{equation}
J(p_{z})=2\pi\int\limits_{|p_{z}|}^{\infty}p\rho(p)dp
\label{electron profile2}
\end{equation}
In this work, we only consider the spherical symmetric cases, and we use the fully relativistic Dirac-Fock approach to calculate the ground states of atomic systems and obtain their Compton profiles.

Given the differential cross sections in Compton scattering, the energy spectrum can be calculated through the integration
\begin{eqnarray}
\bigg(
  \frac{d\sigma}{dT}
\bigg)_{\text{RIA}}
& = &
\int d\Omega_{f}
\bigg(
  \frac{d^{2}\sigma}{d\omega_{f}d\Omega_{f}}
\bigg)_{\text{RIA}} \nonumber
\\
& = &
\int d\Omega_{f}
\frac{r_{0}^{2}}{2}\frac{m_{e}}{q}
\frac{m_{e}c^{2}}{E(p_{z})}
\frac{\omega_{f}}{\omega_{i}}
\overline{X}(p_{z})
J(p_{z}) \nonumber
\\
\label{RIA spectrum}
\end{eqnarray}
with $T=\omega_{i}-\omega_{f}$ to be the energy transfer in Compton scattering. With the atomic many-body effects considered, the RIA formulation overcome the shortcomings in the FEA formulation. Therefore, it is a practical approach to calculate the Compton scattering in the low-energy transfer region or near photoionization threshold region.

Although the RIA formulation effectively takes the atomic bindings, electron pre-collision motions, and other many-body effects into consideration, it still has limitations in dealing with Compton scattering. In RIA formulation, all the many-body effects are incorporated into Compton scattering through the atomic Compton profiles $J(p_{z})$, which is an observable related to the ground state information of atomic systems. The many-body effects coming from the ionized states as well as from the dynamical process of Compton scattering are still lacking. In the past few years, several approaches beyond FEA and RIA have already been investigated \cite{Pratt1,Pratt,Kaplan,Suric,Pratt2,Drukarev1,Drukarev2}. These researches, which mainly employ the low-energy theorems and S-matrix formulation, have revealed many nontrivial properties of Compton scatterings and have attracted lots of interests in interdisciplinary studies. We shall study the atomic Compton Scattering using the more advanced approaches beyond RIA formulation in the future.

\subsection{Dirac-Fock Theory and its Applications \label{sec:2c}}

At the end of this section, we should give a little more discussion on the Dirac-Fock theory, which is a fully relativistic theory served for \emph{ab initio} calculations in the atomic and molecular physics. In this study, the Dirac-Fock theory plays an important role in the Compton scattering calculations by obtaining the ground state wavefunctions and atomic Compton profiles for the atomic systems. The momentum distribution of electrons can be calculated using the Dirac-Fock theory as follows \cite{Desclaux,Qiao}
\begin{equation}
\rho(p) = \sum_{a=1}^{Z}|\phi_{a}(\boldsymbol{p})|^{2}
        = \sum_{njl}N_{njl}
          \bigg(
            (\phi_{njl}^{G}(p))^{2}+(\phi_{njl}^{F}(p))^{2}
          \bigg) \label{rho}
\end{equation}
with $\phi_{njl}^{G}$ and $\phi_{njl}^{F}$ to be the large and small components of electron's momentum wavefunctions of $njl$ subshell. The momentum distribution $\rho(p)$ is further plugged into Eq.(\ref{electron profile2}) to obtain the atomic Compton profiles. This result, combined with Eq.(\ref{RIA spectrum}) in the RIA formulation, gives the energy spectrum of the atomic Compton scattering.

It is worth noting that, apart from the Compton scattering discussed in this study, the relativistic Dirac-Fock theory can be applied to other atomic processes relevant to the dark matter detections. For example, the energy spectrum of electron ionization due to the $\chi-N$ scattering is given by \cite{Moustakidis,Vergados}
\begin{eqnarray}
\frac{d\sigma}{dT}
& = &
\frac{\sigma_{\chi-N}}{2}\sum_{nl}p_{nl}m_{e}q\rho(q)\times  \nonumber
\\
&   &
\bigg\{
  1-\frac{m_{\chi}(T-\varepsilon_{nl})}{\mu_{r}E_{\chi}}+\sqrt{1-\frac{m_{\chi}(T-\varepsilon_{nl})}{\mu_{r}E_{\chi}}}
\bigg\} \nonumber
\\
\end{eqnarray}
Here, $\sigma_{\chi-N}$ is the cross section of $\chi-N$ scattering, $m_{\chi}$ and $E_{\chi}$ are the mass and energy of the dark matter particles, $\mu_{r}$ is the $\chi-N$ reduced mass, $q=\sqrt{2m_{e}T}$ is the momentum transfer, $\varepsilon_{nl}$ is the binding energy for subshell $nl$, and $p_{nl}$ is the probability of finding electron in the subshell $nl$, respectively. In the above equation, $\rho(q)$ is the momentum density of atomic electrons, and it must be calculated using \emph{ab initio} calculations. The fully relativistic Dirac-Fock theory with relativistic effects and more electron correlations incorporated can give more accurate predictions on this process, compared to the previous studies, which mostly calculate $\rho(q)$ using the nonrelativistic Hartree-Fock theory \cite{Moustakidis,Vergados,Ejiri}.

\section{Results and Discussions \label{sec:3}}

\begin{figure}
  \centering
  \includegraphics[width=0.475\textwidth]{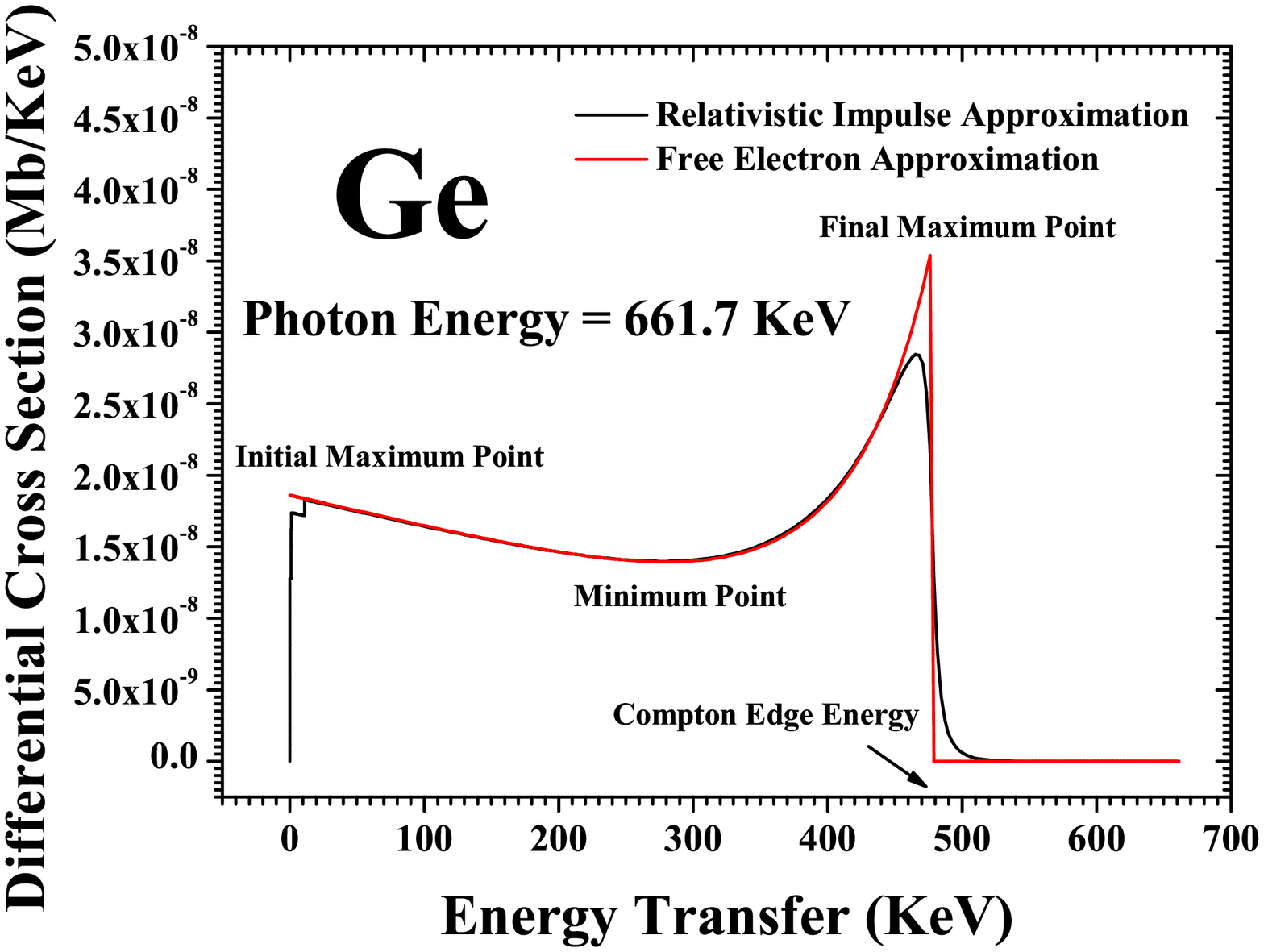}
  \includegraphics[width=0.475\textwidth]{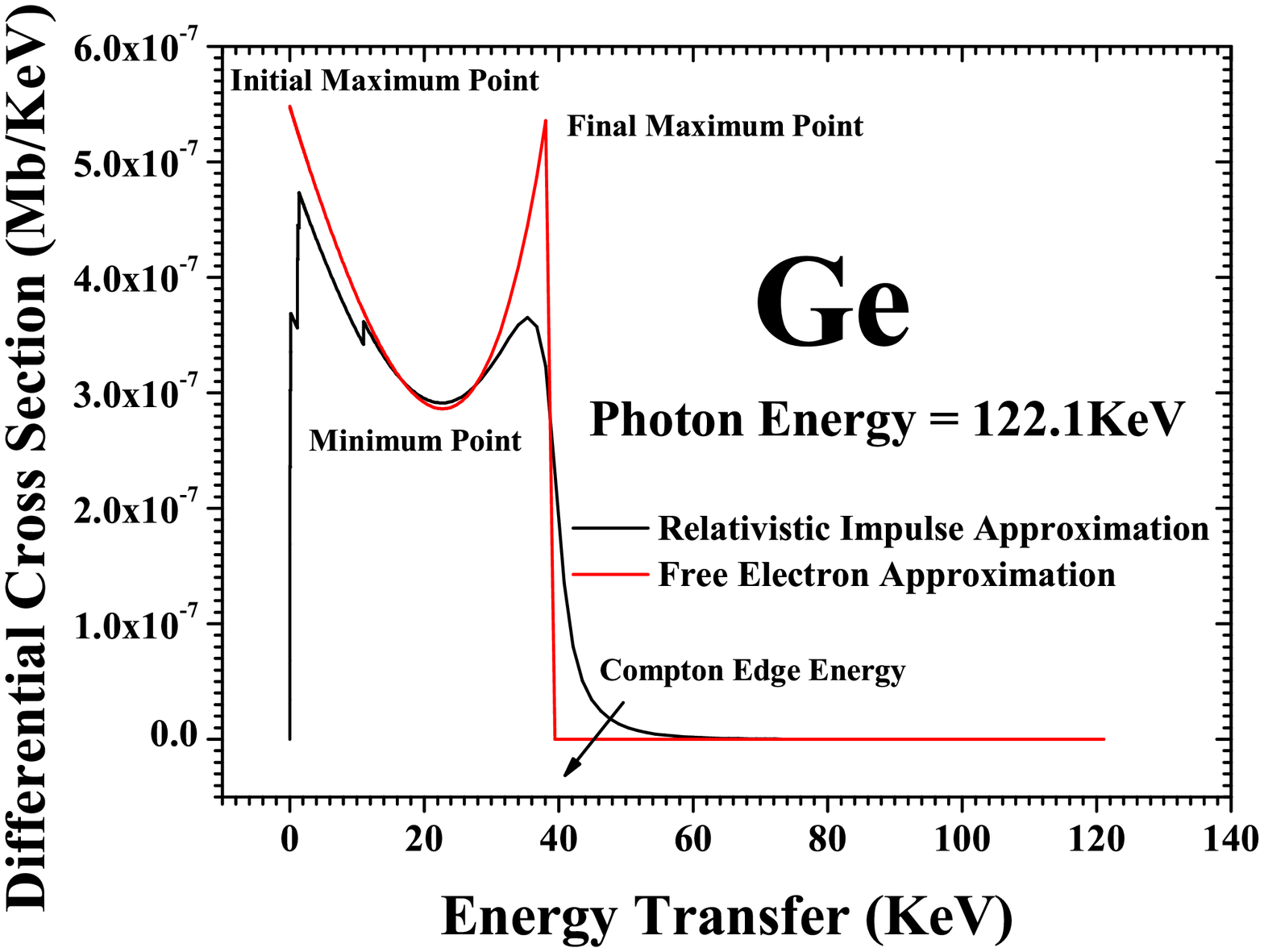}
  \caption{The Compton scattering energy spectrum in the free electron approximation (FEA) and relativistic impulse approximation (RIA). We choose the energy spectrum for Ge atom at incident photon energies $661.7$ KeV and $122.1$ KeV as examples to illustrate the differences between the FEA and RIA results. The initial maximum point, final maximum point, and minimum point are labeled in this figure. The locations of the Compton edge in the FEA results are also presented in this figure.}
\label{energy spectrum}
\end{figure}

In this section, we present our results on the Compton scattering energy spectrum obtained from FEA and RIA formulations. In direct detection experiments of dark matter particles, both the dynamical properties of atomic Compton scattering and the geometrical structure of detectors can affect the observed Compton scattering energy spectrum in a real experiment. For simplicity, in this work we only focus ourselves to the dynamical properties of Compton scattering processes. The effects coming from the geometrical structure of detectors are leaving for future studies. In the present work, we consider the simple Si and Ge atomic systems, which correspond to the ideal point-like Si and Ge detectors. The ground states of Si and Ge atoms are calculated within the fully relativistic Dirac-Fock theory \cite{Desclaux,Grant}.

The Compton scattering energy spectra calculated from FEA and RIA formulations are shown in Fig. \ref{energy spectrum}.  We choose the energy spectrum for Ge atom at incident photon energies $661.7$ KeV and $122.1$ KeV as examples to illustrate the differences between the FEA and RIA results. In the FEA result, the energy transfer has a maximum, and the maximal energy transfer case corresponds to the Compton edge. However, in RIA formulation, the energy transfer can exceed the Compton edge due to the atomic many-body effects. We study various aspects of Compton scattering energy spectrum in subsection \ref{sec:3a}-\ref{sec:3c}, and comparisons between our theoretical calculations and Monte Carlo simulations are presented in subsection \ref{sec:3d}.

To study the Compton scattering energy spectrum in low and high energy regions, throughout this work, we pick several characteristic gamma-ray energies $122.1$ KeV, $238.6$ KeV, $356.0$ KeV, $511.0$ KeV, $661.7$ KeV, $1173.2$ KeV, $1332.5$ KeV, and $1460.7$ KeV as examples to show our results. These gamma-rays can be emitted from the $^{57}Co$ ($122.1$ KeV), $^{212}Pb$ ($238.6$ KeV), $^{133}Ba$ ($356.0$ KeV), $^{22}Na$ ($511.0$ KeV), $^{137}Cs$ ($661.7$ KeV), $^{60}Co$ ($1173.2$ KeV, $1332.5$ KeV) and $^{40}K$ ($1460.7$ KeV) characteristic gamma-ray sources, respectively.

\subsection{Ratios between Maximum and Minimum \label{sec:3a}}

In the Fig. \ref{energy spectrum}, we have labeled some important points in the Compton scattering energy spectrum for FEA and RIA formulations. They are initial maximum point, final maximum point, and minimum point, respectively. In the FEA formulation, the initial maximum point is obtained by taking the $T\rightarrow 0$ limit in the energy spectrum. However, in the RIA formulation, the differential cross section $d\sigma/dT \rightarrow 0$ when $T\rightarrow 0$, thus the initial maximum point is given by the local maximum of $d\sigma/dT$ in the low-energy transfer region. In most cases, this initial maximum point locates at energy just above the K ionization threshold, except for the case of low energy incoming photon $\omega_{i}=122.1$ KeV for Ge atom. In the Compton scattering energy spectrum, the locations of these maximum and minimum points can be shifted when the incident photon energy varies. Moreover, the ratios between maximum and minimum of $d\sigma/dT$ in energy spectrum are also influenced by incoming photon energy.

\begin{figure}
  \centering
  \includegraphics[width=0.475\textwidth]{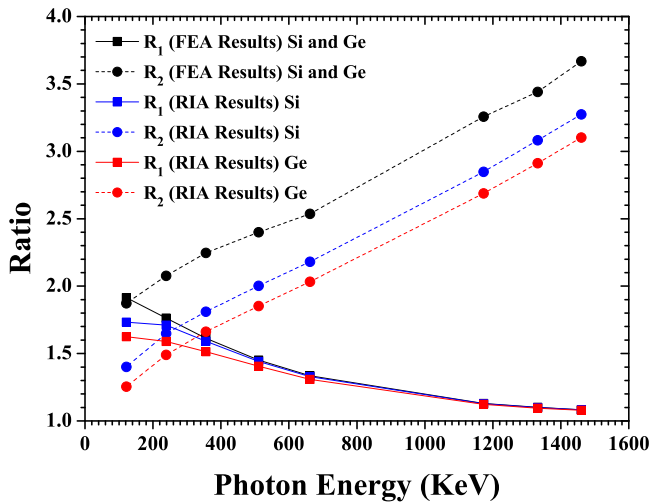}
  \caption{The ratios between maximum and minimum of $d\sigma/dT$ in the Compton scattering energy spectrum from FEA and RIA calculations for Si and Ge atoms.  The two parameters $R_{1}$ and $R_{2}$ are defined in Eqs. (\ref{R1})-(\ref{R2}), respectively.}
\label{height ratio}
\end{figure}

In order to compare the ratio between maximum and minimum in Compton scattering energy spectrum quantitatively, we can define the two parameters as follows
\begin{eqnarray}
R_{1} & \equiv & \frac{(d\sigma/dT)_{\text{imax}}}{(d\sigma/dT)_{\text{min}}} \label{R1}
\\
R_{2} & \equiv & \frac{(d\sigma/dT)_{\text{fmax}}}{(d\sigma/dT)_{\text{min}}} \label{R2}
\end{eqnarray}
with $(d\sigma/dT)_{\text{imax}}$, $(d\sigma/dT)_{\text{fmax}}$, and $(d\sigma/dT)_{\text{min}}$ to be the singly differential cross section at initial maximum point, final maximum point, and minimum point, respectively.

The calculation results of $R_{1}$ and $R_{2}$ are plotted in Fig. \ref{height ratio} for different photon energies in FEA and RIA formulation for Si and Ge atoms. In FEA formulation, the parameters $R_{1}$ and $R_{2}$ are the same for all elements. While in RIA formulation, $R_{1}$ and $R_{2}$ depend on the target materials. From this figure, it is apparent that the parameter $R_{1}$ decreases when photon energy becomes higher, while the parameter $R_{2}$ increases when photon energy increases. The same trend appears for both FEA and RIA results. This result shows that in the Compton scattering, the ``initial'' of energy spectrum becomes lower and the ``tail'' of energy spectrum becomes higher with the increase of incident photon energy.

Furthermore, in Fig. \ref{height ratio}, we can see various curves of $R_{1}$ parameter converge to each other when photon energy exceeds 1 MeV, but the result of $R_{2}$ parameter does not present this convergence. Approximately, the $R_{2}$ parameter exhibits linear structures with respect to the incident photon energy, both in FEA and RIA results. Apparently, Fig. \ref{height ratio} indicates that the FEA results successfully predict the $R_{1}$ parameter, except for the cases of low energy photons, in which the location of initial maximum point in RIA calculations can be significantly shifted due to the photoionization threshold, as presented in Fig. \ref{energy spectrum}. The FEA results fail to predict the $R_{2}$ parameter because the atomic binding effects can smear out the Compton edge and then reduces the peak, and the extent of smearing out is influenced by the target materials. However, the simple FEA result can predict the slope of the $R_{2}$ parameter with respect to the incident photon energy.

\subsection{Linear Fit method and Slopes \label{sec:3b}}

\begin{figure}
  \centering
  \includegraphics[width=0.475\textwidth]{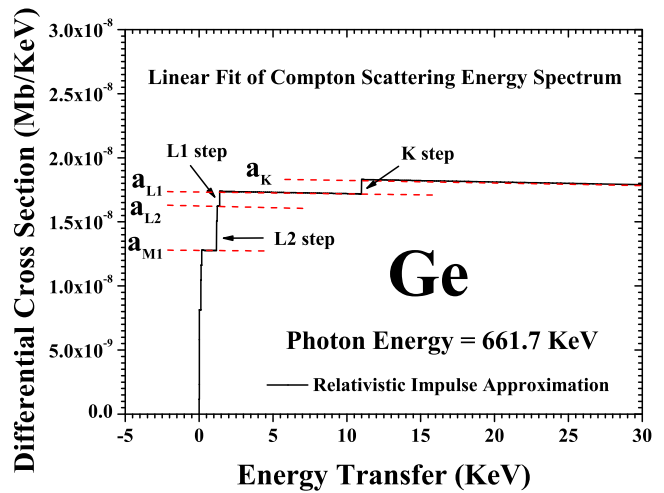}
  \caption{The linear fit method of Compton scattering energy spectrum as a diagrammatic sketch. The linear platforms and their corresponding slopes of the K, L1, L2, and M1 subshells are labeled in this figure. The step structures at K, L1, and L2 photoionization thresholds are also indicated in the figure.}
\label{linear fit}
\end{figure}

\begin{figure}
  \centering
  \includegraphics[width=0.475\textwidth]{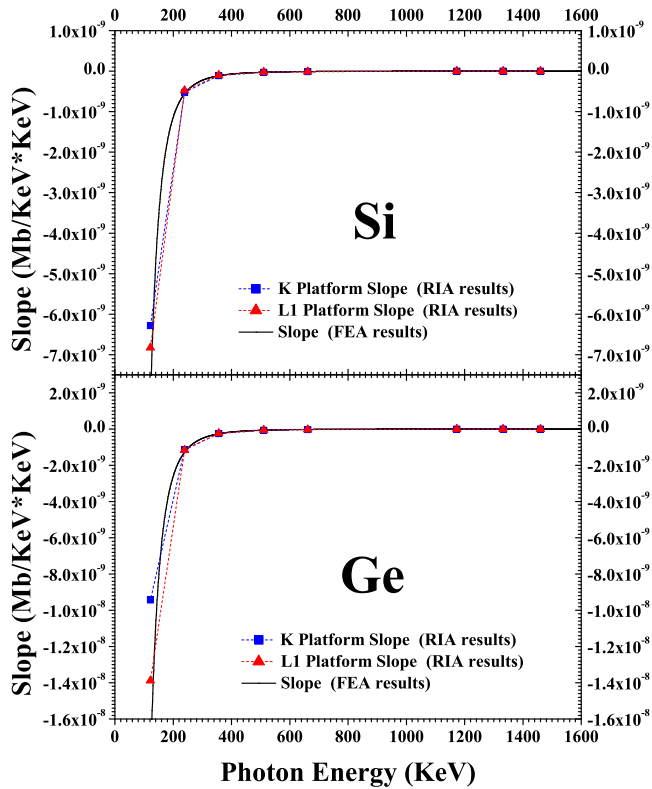}
  \caption{The slopes of K and L1 platforms in the Compton scattering energy spectrum obtained from the RIA calculations for Si and Ge atoms. The slopes of Si and Ge atoms obtained from the FEA are also plotted in this figure as comparisons.}
\label{slope}
\end{figure}

In the low-energy transfer region, the Compton scattering energy spectrum exhibits several linear platforms and steps structures  approximately. This linearity is especially noticeable when the energy transfer $T$ is close to the photoionization thresholds, which are $1.8$ KeV and $11.1$ KeV for Si and Ge atoms, respectively. This allows us to employ a linear fit method for Compton scattering energy spectrum in the near photoionization threshold region, which gives
\begin{equation}
\frac{d\sigma}{dT} \approx aT+b
\end{equation}
In the FEA calculations, we get an overall slope $a^{\text{FEA}}$ for the whole energy spectrum. In the RIA calculations, for various subshells we can get their corresponding platforms and slopes $a^{\text{RIA}}_{\text{K}}$, $a^{\text{RIA}}_{\text{L1}}$, $a^{\text{RIA}}_{\text{L2}}$,$a^{\text{RIA}}_{\text{M1}}$ respectively. Furthermore, in the FEA results, the slope $a^{\text{FEA}}$ is inversely proportional to the quartic power of incoming photon energy, which can be shown directly from the analytical expressions of FEA. In the low-energy transfer limit $T = \omega_{i}-\omega_{C} \rightarrow 0$, the Compton scattering energy spectrum in FEA formulation becomes
\begin{eqnarray}
\bigg(
  \frac{d\sigma}{dT}
\bigg)_{\text{FEA}}
& = &
\pi r_{0}^{2} \frac{m_{e}c^{2}}{\omega_{i}^{2}}
\bigg[
  \frac{\omega_{i}}{\omega_{C}}
  +\frac{\omega_{C}}{\omega_{i}}
  -2m_{e}c^{2}
    \bigg( \frac{1}{\omega_{C}}-\frac{1}{\omega_{i}} \bigg) \nonumber
\\
&   &
  +m_{e}^{2}c^{4}
    \bigg( \frac{1}{\omega_{C}}-\frac{1}{\omega_{i}} \bigg)^{2}
\bigg] \nonumber
\\
& \approx &
2\pi r_{0}^{2} \frac{m_{e}c^{2}}{\omega_{i}^{2}}
-2\pi r_{0}^{2} \frac{m_{e}^{2}c^{4}}{\omega_{i}^{4}} T
+o(T^{2}) \nonumber
\\
& \approx &
a^{\text{FEA}}T+b^{\text{FEA}}+o(T^{2})
\end{eqnarray}
which gives $a^{\text{FEA}}=-2\pi r_{0}^{2}m_{e}^{2}c^{4}/\omega_{i}^{4}$. The linear fit method is illustrated as a diagrammatic sketch in the Fig. \ref{linear fit} for RIA calculations of Ge atom with incident photon energy $\omega_{i}=661.7$ KeV.

The results of the slopes of K and L1 platforms $a^{\text{RIA}}_{\text{K}}$, $a^{\text{RIA}}_{\text{L1}}$ in RIA calculations are presented in the Fig. \ref{slope} for Si and Ge atom for various incident photon energies. The slopes of energy spectrum obtained from FEA calculations $a^{\text{FEA}}$ are also plotted in this figure as comparisons. From the Fig. \ref{slope}, we can observe that all slopes are negative for the FEA and RIA results, and each slope diminishes as photon energy increases. Therefore, we can draw the conclusion that the Compton scattering energy spectrum becomes steeper for lower energy incoming photons, and the energy spectrum is comparatively flatter for higher energy gamma-rays. Moreover, the slopes of K and L1 platforms in RIA calculations converge to each other for Si and Ge atoms, and they are consistent with the FEA results, except for low-energy $122.1$ KeV photon. This indicate that, for high incident photon energies, the electrons in K and L shells almost contribute equivalently as free electrons in the Compton scattering energy spectrum.

\subsection{Altitudes of Platforms \label{sec:3c}}

\begin{figure}
  \centering
  \includegraphics[width=0.475\textwidth]{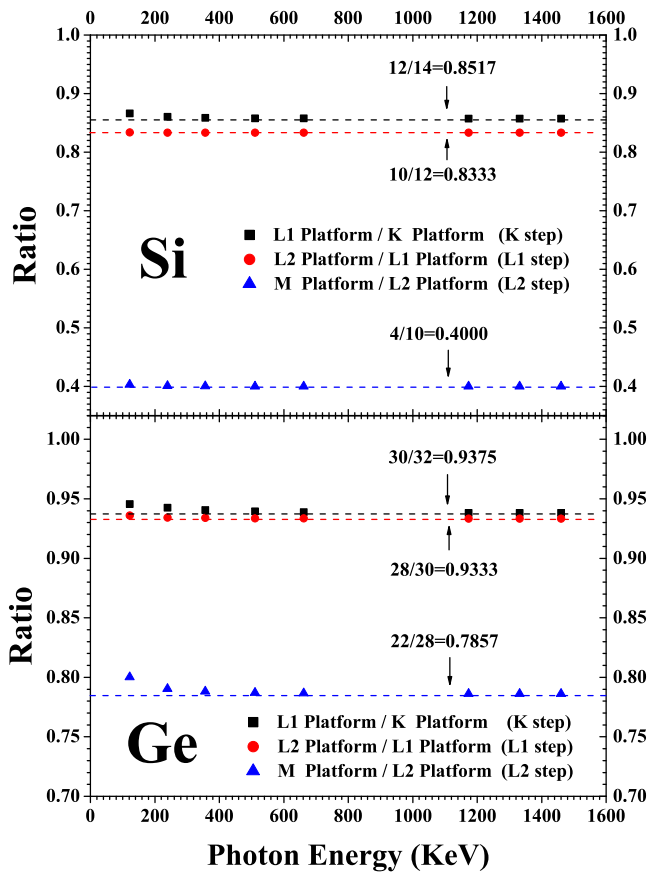}
  \caption{The relative altitude ratios between platforms on two sides of K, L1, and L2 steps at their corresponding photoionization thresholds in RIA calculations for Si and Ge atoms. For convenience, several lines are added in this figure to illustrate the ratios of electron's numbers that activated in Compton scattering on two sides of K, L1, and L2 steps.}
\label{shell ratio}
\end{figure}

After employing the linear fit method displayed in Fig. \ref{linear fit} for the Compton scattering energy spectrum in the low-energy transfer region, we can get the linear platform for each subshell. These platforms, when combined together, present step structures in the energy spectrum. The altitudes of platforms for different subshells are influenced by the incident photon energy, and they need to be studied quantitatively. In this subsection, we analyze the relative altitude ratios between the platforms on two sides of the K, L1 and L2 steps at their photoionization thresholds.

\begin{table*}
\caption{Numbers of electrons in different subshells for Si and Ge atoms. In this table, we give the atomic subshell labels in the nonrelativistic and relativistic theories as well as the number of electron (N. E.) in each subshell.}
\label{table}
\centering
\vspace{2mm}
\begin{ruledtabular}
\begin{tabular}{lcccccccccccc}
\multicolumn{10}{c}{Si}
\\
\hline
Label        & K  & L1 & L2         & M1 & M2         &            &    &            & Total
\\
\hline
Subshell(NR) & 1s & 2s & 2p         & 3s & 3p         &            &    &            &
\\
Subshell(R)  & 1s & 2s & 2p$_{1/2}$ & 3s & 3p$_{1/2}$ &            &    &            &
\\
             &    &    & 2p$_{3/2}$ &    & 3p$_{3/2}$ &            &    &            &
\\
\hline
N. E.        & 2  & 2  & 6          & 2  & 2          &            &    &            & 14
\\
\hline
\multicolumn{10}{c}{Ge}
\\
\hline
Label        & K  & L1 & L2         & M1 & M2         & M3         & N1 & N2         & Total
\\
\hline
Subshell(NR) & 1s & 2s & 2p         & 3s & 3p         & 3d         & 4s & 4p         &
\\
Subshell(R)  & 1s & 2s & 2p$_{1/2}$ & 3s & 3p$_{1/2}$ & 3d$_{3/2}$ & 4s & 4p$_{1/2}$ &
\\
             &    &    & 2p$_{3/2}$ &    & 3p$_{3/2}$ & 3d$_{5/2}$ &    & 4p$_{3/2}$ &
\\
\hline
N. E.        & 2  & 2  & 6          & 2  & 6          & 10         & 2  & 2          & 32
\\
\end{tabular}
\end{ruledtabular}
\end{table*}

The results of the relative altitude ratios between the platforms on two sides of K, L1, and L2 steps are given for RIA calculations in Fig. \ref{shell ratio} for Si and Ge atoms. For convenience, we add several lines in this figure to illustrate the ratio of electron's number that activated in the Compton scattering on both sides of K, L1, and L2 steps. If electrons in different subshells contribute equivalently in the Compton scattering energy spectrum, then the relative altitude ratio of platforms will be the ratio of electrons activated in Compton scattering. We shall take the K step for Si atom as an example to explain this point. When energy transfer is larger than K shell photoionization threshold, all electrons are activated in the Compton scattering. However, once the energy transfer fall below the photoionization threshold of K shell, two $1s$ electrons become inactivated in the scattering process. If electrons in different subshells contribute equivalently in Compton scattering, then the relative altitude ratio between platforms on two sides of K step for the Si atom will be $12/14=0.8333$. Results of other situations can be analyzed in a similar way. For the reader's convenience, the numbers of electrons in different subshells are given for Si and Ge atoms in the table \ref{table}. From this figure, we can learn that the relative altitude ratios between platforms on two sides of K, L1, and L2 steps almost fall on these ideal lines for Si and Ge atoms, except for low-energy photons for Ge atom. Therefore, we can draw the conclusion that K and L shell electrons almost contribute equivalently in the Compton scattering energy spectrum, except for low incident photon energies. This is consistent with our results obtained in subsection \ref{sec:3b}.

\subsection{Comparison of Our Results with Monte Carlo Simulations \label{sec:3d}}

In the present work, we also compare our results with the Monte Carlo simulations from Peng Gu \emph{et al.} for the relative altitude ratio between platforms on two sides of K step of Ge detectors \cite{Gu}. The conventional Monte Carlo simulation program Geant4, which has been extensively applied to nuclear physics, astrophysics, and elementary particle physics, is adopted in Gu's simulations \cite{Geant4,Geant4b}. In simulations, the same linear fit approach is used to extract information of relative altitude ratio between platforms on two sides of K step. It is worth while to note that we use the simple atomic model in the theoretical calculations, and the geometrical structure of detectors are ignored. Thus, our theoretical results only correspond to the cases of ideal point-like detectors, as mentioned in the beginning of Section \ref{sec:3}. While in Gu's simulations, the geometrical structures of detectors are takin into consideration. The detectors' geometrical structures are chosen exactly the same as the high-purity Ge detectors used in CDEX-10 and Super-CDMS experiments of dark matter direct detections \cite{CDMS,CDEX}. Two simulative models in Geant4 program (version 10.3) are employed to simulate the Compton scattering processes. They are Livermore and Monash models, respectively, and both of them utilize the ideas of RIA to deal with the atomic Compton scattering \cite{Geant4b,Livermore,Monash}.

\begin{figure}
  \centering
  \includegraphics[width=0.475\textwidth]{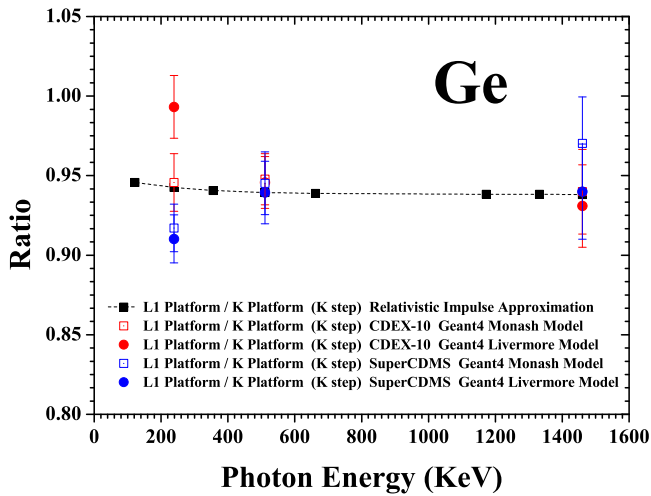}
  \caption{The comparative results between theoretical calculations and Monte Carlo simulations on the relative altitude ratio between platforms on both sides of K step for Ge detectors. For simulative results, the standard deviations are presented in this figure.}
\label{shell ratio simulation}
\end{figure}

The comparative results between theoretical calculations and Monte Carlo simulations are shown in Fig. \ref{shell ratio simulation}. For simulative results, the relative altitude ratio between platforms on two sides of K step for CDEX-10 and Super-CDMS detectors obtained using different simulation models are presented, and the standard deviations are also given in this figure. However, in our theoretical calculations, the linear fit method works perfect well and the standard deviations are less than 1\%. Therefore, the standard deviations of the relative altitude ratios in our calculations are not given in this figure. Fig. \ref{shell ratio simulation} indicates that there are obvious deviations between simulative results obtained using CDEX-10 and Super-CDMS Ge detectors. Therefore, it is clearly that the geometrical structures of detectors indeed affect the relative altitude ratio between platforms on two sides of K step for Ge atom in the Compton scattering energy spectrum. Throughout the above comparisons, we can see that the theoretical and simulative results are comparatively consistent with each other, with certain deviations possibly arising from the influence of detector's geometrical structure. An experiment aiming to test our theoretical calculations as well as the Monte Carlo simulations is ongoing now \cite{Gu,Experiment}.

\section{Summary and Conclusions \label{sec:4}}

Throughout this study, we investigate the Compton scattering energy spectrum for Si and Ge atoms in FEA and RIA formulations. The ratios between maximum and minimum in energy spectrum are analyzed quantitatively. In the low-energy transfer or near photoionization threshold regions, the Compton scattering energy spectrum presents approximately linear platform and step structure for each subshell. We use a linear fit method for energy spectrum to extract slopes for K and L1 platforms in RIA calculations. The relative altitude ratios between platforms on two sides of K and L steps at their corresponding photoionization thresholds are also analyzed in RIA calculations.

Our results show that, in FEA and RIA calculations of Compton scattering, the ``initial'' of energy spectrum goes higher and the ``tail'' of energy spectrum becomes lower when the incoming photon energy reduces. Particularly, in Compton scattering energy spectrum the ratio between final maximum and minimum values, which is characterized by parameter $R_{2}$ in subsection \ref{sec:3a}, appear approximately linear relationship with respect to the incident photon energy. Through analysis of slopes and step structures for various subshells, we can draw the conclusion that the energy spectrum becomes steeper for lower energy gamma-rays, and the spectrum is comparatively flatter for high energy photons. The electrons from K and L shell almost contribute equivalently in the Compton scattering energy spectrum, except for the cases of low energy incident photons.

Furthermore, we also compare our theoretical results with the Monte Carlo simulations, and comparatively consistent results are obtained with certain deviations possibly arising from the influence of detector's geometrical structure. The theoretical and simulative results can be test by experiments ongoing. We wish that our work could be useful in the X-ray and gamma-ray background understanding, background analysis and background subtraction in dark matter direct detection experiments, especially in the low-energy transfer or near photoionization threshold regions.

\section*{ACKNOWLEDGMENTS}

We acknowledge helpful discussions with Hao-Yang Xin, Li-Tao Yang, Yu-Feng Zhou, Keh-Ning Huang and Herry T. Wong. This work was supported by the National Key Research and Development Program of China (Grant No. 2017YFA0402203), the National Natural Science Foundation of China (Grant No. 11475117, No. 11975159 and No. 11975162) and the Fundamental Research Funds for the Central Universities.


\begin{thebibliography}{99}

\bibitem{Cooper1}
    M. J. Cooper, \emph{Compton scattering and electron momentum distributions}, Adv. Phys. {\bf 20}, 453-491 (1971).

\bibitem{Cooper2}
    M. J. Cooper, \emph{Compton scattering and electron momentum determination}, Rep. Prog. Phys. {\bf 48(4)}, 415-481 (1985).

\bibitem{Cooper3}
    M. J. Cooper, \emph{Compton scattering and the study of electron momentum density distributions}, Radiat. Phys. Chem. {\bf 50}, 63-76 (1997).

\bibitem{Ji}
    X. Ji, \emph{Deeply virtual Compton scattering}, Phys. Rev. D {\bf 55}, 7114-7125 (1997).

\bibitem{Porter}
    T. A. Porter, I. V. Moskalenko, A. W. Strong, E. Orlando and L. Bouchet, \emph{Inverse Compton Origin of the Hard X-Ray and Soft Gamma-Ray Emission from the Galactic Ridge}, Astrophys. J {\bf 682}, 400-407 (2008).

\bibitem{Wang}
    Y. J. Wang, H. Lin, B. Barbiellini, P. E. Mijnarends, S. Kaprzyk, R. S. Markiewicz and A. Bansil, \emph{Proposal to determine the Fermi-surface topology of a doped iron-based superconductor using bulk-sensitive Fourier-transform Compton scattering}, Phys. Rev. B {\bf 81}, 092501 (2010).

\bibitem{Phuoc}
    K. T. Phuoc, S. Corde, C. Thaury, V. Malka, A. Tafzi, J. P. Goddet, R. C. Shah, S. Sebban and A. Rousse, \emph{All-optical Compton gamma-ray source}, Nat. Photonics {\bf 6}, 308-311 (2012).

\bibitem{Takada}
    A. Takada \emph{et al.}, \emph{Development of an advanced Compton camera with gaseous TPC and scintillator}, Nucl. Insrum. Meth. A {\bf 546}, 258-262 (2005).

\bibitem{Mihailescu}
    L. Mihailescu, K. M. Vetter, M. T. Burks, E. L. Hull and W. W. Craig, \emph{SPEIR: A Ge Compton camera}, Nucl. Insrum. Meth. A {\bf 570}, 89-100 (2007).

\bibitem{Chiu}
    J. -L. Chiu \emph{et al.}, \emph{The upcoming balloon campaign of the Compton Spectrometer and Imager (COSI)}, Nucl. Insrum. Meth. A {\bf 784}, 359-363 (2015).

\bibitem{Zwicky}
    F. Zwicky, \emph{Die Rotverschiebung von extragalaktischen Nebeln}, Helv. Phys. Acta {\bf 6}, 110-127 (1933).

\bibitem{Rubin}
    V. C. Rubin and W. K. Ford, \emph{Rotation of the Andromeda Nebula from a Spectroscopic Survey of Emission Regions}, Astrophys. J. {\bf 159}, 379-403 (1970).

\bibitem{Corbelli}
    E. Corbelli and P. Salucci, \emph{The extended rotation curve and the dark matter halo of M33}, Mon. Not. R. Astron. Soc. {\bf 311}, 441-447 (2000).

\bibitem{Clowe}
    D. Clowe, M. Bradac, A. H. Gonzalez, M. Markevitch, S. W. Randall, C. Jones and D. Zaritsky, \emph{A Direct Empirical Proof of the Existence of Dark Matter}, Astrophys. J. {\bf 648}, L109-L113 (2006).

\bibitem{Blumenthal}
    G. R. Blumenthal, S. M. Faber, J. R. Primack and M. J. Rees, \emph{Formation of galaxies and large-scale structure with cold dark matter}, Nature {\bf 311}, 517-525 (1984).

\bibitem{Davis}
    M. Davis, G. Efstathiou, C. S. Frenk and S. D. M. White, \emph{The evolution of large-scale structure in a universe dominated by cold dark matter}, Astrophys. J. {\bf 292}, 371-394 (1985).

\bibitem{WMAP}
    E. Komatsu \emph{et al.} (WMAP Collaboration), \emph{Seven-Year Wilkinson Microwave Anisotropy Probe (WMAP) Observations: Cosmological Interpretation},  Astrophys. J. Suppl. {\bf 192}, 18 (2011).

\bibitem{Planck}
    P. A. R. Ade \emph{et al.} (Planck Collaboration), \emph{Planck 2013 results. XVI. Cosmological parameters}, Astron. Astrophys. {\bf 571}, A16 (2014).

\bibitem{Undagoitia}
    T. M. Undagoitia and L. Rauch, \emph{Dark matter direct detection experiments}, J. Phys. G: Nucl. Part. Phys. {\bf 43}, 013001 (2015).

\bibitem{Bertone}
    G. Bertone, D. Hooper and J. Silk, \emph{Particle dark matter: Evidence, candidates and constraints}, Phys. Rept. {\bf 405}, 279-390 (2005).

\bibitem{RDG2016}
    C. Patrignani \emph{et al.} (Particle Data Group), \emph{Review of Particle Physics}, Chin. Phys. C {\bf 40}, 100001 (2016).

\bibitem{RDG2018}
    M. Tanabashi \emph{et al.} (Particle Data Group), \emph{Review of Particle Physics}, Phys. Rev. D {\bf 98}, 030001 (2018).

\bibitem{Barker}
    D. Barker (on behalf of the SuperCDMS Collaboration), \emph{Low Energy Background Spectrum in CDMSlite}, Proceedings of Science {\bf 282}, 874 (2017).

\bibitem{Ramanathan}
    K. Ramanathan, A. Kavner, A. E. Chavarria, P. Privitera, D. Amidei, T.-L. Chou, A. Matalon, R. Thomas, J. Estrada, J. Tiffenberg and J. Molina, \emph{Measurement of low energy ionization signals from Compton scattering in a charge-coupled device dark matter detector}, Phys. Rev. D {\bf 96}, 042002 (2017).

\bibitem{Robinson}
    A. E. Robinson, \emph{Coherent photon scattering background in sub-GeV$/c^{2}$ direct dark matter searches}, Phys. Rev. D {\bf 95}, 021301(R) (2017); Erratum: Phys. Rev. D {\bf 95}, 069907 (2017)

\bibitem{Bernabei}
    R. Bernabei \emph{et al.}, \emph{Investigating electron interacting dark matter}, Phys. Rev. D {\bf 88}, 023506 (2008).

\bibitem{Agnes}
    P. Agnes \emph{et al.}, \emph{Constraints on Sub-GeV Dark-Matter-Electron Scattering from the DarkSide-50 Experiment}, Phys. Rev. Lett. {\bf 121}, 111303 (2018).

\bibitem{Moustakidis}
    C. C. Moustakidis, J. D. Vergados and H. Ejiri, \emph{Direct dark matter search by observing electrons produced in neutralino-nucleus collisions}, Nucl. Phys. B {\bf 727}, 406-420 (2005).

\bibitem{Vergados}
    J. D. Vergados and H. Ejiri, \emph{The role of ionization electrons in direct neutralino detection}, Phys. Lett. B {\bf 606}, 313-322 (2005).

\bibitem{Ejiri}
    H. Ejiri, C. C. Moustakidis and J. D. Vergados, \emph{Dark matter search by exclusive studies of X-rays following WIMPs nuclear interactions}, Phys. Lett. B {\bf 639}, 218-222 (2006).

\bibitem{Ibe}
    M. Ibe, W. Nakano, Y. Shojia and K. Suzuki, \emph{Migdal effect in dark matter direct detection experiments}, JHEP {\bf 2018(03)}, 194 (2018).

\bibitem{Klein-Nishina}
    O. Klein and Y. Nishina, \emph{\"Uber die Streuung von Strahlung durch freie Elektronen nach der neuen relativistischen Quantendynamik von Dirac}, Z. Phys. {\bf 52},  853-868 (1929).

\bibitem{Sakurai}
    J. J. Sakurai, \emph{Advanced Quantum Mechanics} (Addison-Wesley, New-York, 1967).

\bibitem{Eisenberger1}
    P. Eisenberger and P. M. Platzman, \emph{Compton Scattering of X Rays from Bound Electrons}, Phys. Rev. A {\bf 2}, 415-423 (1970).

\bibitem{Eisenberger2}
    P. Eisenberger and W. A. Reed, \emph{Relationship of the relativistic Compton cross section to the electron's velocity distribution}, Phys. Rev. B {\bf 9}, 3237-3241 (1974).

\bibitem{Ribberfors1}
    R. Ribberfors, \emph{Relationship of the relativistic Compton cross section to the momentum distribution of bound electron states}, Phys. Rev. B {\bf 12}, 2067-2074 (1975).

\bibitem{Ribberfors2}
    R. Ribberfors, \emph{Relationship of the relativistic Compton cross section to the momentum distribution of bound electron states--II. Effects of anisotropy and polarization}, Phys. Rev. B {\bf 12}, 3136-3141 (1975).

\bibitem{Ribberfors3}
    R. Ribberfors and K.-F. Berggren, \emph{Incoherent-x-ray-scattering functions and cross sections $(d\sigma/d\Omega)_{incoh}$ by means of a pocket calculator}, Phys. Rev. A {\bf 26(6)}, 3325-3333 (1982).

\bibitem{Ribberfors4}
    R. Ribberfors,  \emph{X-ray incoherent scattering total cross sections and energy-absorption cross sections by means of simple calculation routines}, Phys. Rev. A {\bf 27}, 3061-3070 (1983); Erratum: Phys. Rev, A {\bf 28}, 2551 (1983).

\bibitem{Kubo}
    Y. Kubo, \emph{Electron correlation effects on Compton profiles of copper in the GW approximation}, J. Phys. Chem. Solids {\bf 66} 2202-2206 (2005).

\bibitem{Rathor}
    A. Rathor, V. Sharma, N. L. Heda, Y. Sharma and B. L. Ahuja, \emph{Compton profiles and band structure calculations of IV-VI layered compounds GeS and GeSe}, Radiat. Phys. Chem. {\bf 77}, 391-400 (2008).

\bibitem{Pisani}
    C. Pisani, M. Itou, Y. Sakurai, R. Yamaki, M. Ito, A. Erba and L. Maschio, \emph{Evidence of instantaneous electron correlation from Compton profiles of crystalline silicon}, Phys. Chem. Chem. Phys. {\bf 13}, 933-936 (2011).

\bibitem{Aguiar}
    J. C. Aguiar, D. Mitnik and H. O. Di Rocco, \emph{Electron momentum density and Compton profile by a semi-empirical approach}, J. Phys. Chem. Solids {\bf 83}, 64-69 (2015).

\bibitem{Brusa}
    D. Brusa, G. Stutz, J. A. Riveros, J. M. Fern\'andez-Varea and F. Salvat, \emph{Fast sampling algorithm for the simulation of photon Compton scattering}, Nucl. Insrum. Meth. A {\bf 379}, 167-175 (1996).

\bibitem{Geant4}
    J. Allison \emph{et al.}, \emph{Recent developments in GEANT4}, Nucl. Insrum. Meth. A {\bf 835}, 186-225 (2016)

\bibitem{Geant4b}
    GEANT Collaboration, \emph{GEANT4 Physics Reference Manual, Version 10.3}, Available online at http://GEANT4.web.cern.ch/ (2016).

\bibitem{Livermore}
    D. E. Cullen, \emph{A simple model of photon transport}, Nucl. Insrum. Meth. B {\bf 101}, 499-510 (1995).

\bibitem{Monash}
    J. M. C. Brown, M. R. Dimmock, J. E. Gillam and D. M. Paganin, \emph{A low energy bound atomic electron Compton scattering model for Geant4}, Nucl. Insrum. Meth. B {\bf 338}, 77-88 (2014).

\bibitem{Biggs}
    F. Biggs, L. B. Mendelsohn and J. B. Mann, \emph{Hartree-Fock Compton Profiles for the Elements}, At. Data and Nucl. Data Table {\bf 16}, 201-309 (1975).

\bibitem{Hubbell}
    J. H. Hubbell, W. J. Veigele, E. A. Briggs, R. T. Brown, D. T. Cromer and R. J. Howerton, \emph{Atomic form factors, incoherent scattering functions, and photon scattering cross sections}, J. Phys. Chem. Ref. Data {\bf 4}, 471-538 (1975); Erratum: J. Phys. Chem. Ref. Data {\bf 6}, 615 (1977).

\bibitem{EPDL}
    D. E. Cullen, J. H. Hubbell and L. Kissel, \emph{EPDL97: the evaluated photon data library}, Lawrence Livermore National Laboratory Report UCRL-50400 {\bf 6(5)}, 1-28 (1997).

\bibitem{CoGeNT}
    C. E. Aalseth \emph{et al.} (CoGeNT Collaboration), \emph{CoGeNT: A search for low-mass dark matter using p-type point contact germanium detectors}, Phys. Rev. D {\bf 88}, 012002 (2013).

\bibitem{CDMS}
    R. Agnese \emph{et al.} (SuperCDMS Collaboration), \emph{New Results from the Search for Low-Mass Weakly Interacting Massive Particles with the CDMS Low Ionization Threshold Experiment}, Phys. Rev. Lett. {\bf 116}, 071301 (2016)

\bibitem{CDEX}
    H. Jiang \emph{et al.} (CDEX Collabration), \emph{Limits on Light Weakly Interacting Massive Particles from the First 102.8 kg $\times$ day Data of the CDEX-10 Experiment}, Phys. Rev. Lett. {\bf 120}, 241301 (2018).

\bibitem{CDMS2}
    R. Agnese \emph{et al.} (SuperCDMS Collaboration), \emph{Results from the Super Cryogenic Dark Matter Search Experiment at Soudan}, Phys. Rev. Lett. {\bf 120}, 061802 (2018).

\bibitem{Monroe}
    J. Monroe and P. Fisher, \emph{Neutrino backgrounds to dark matter searches}, Phys. Rev. D {\bf 77}, 033007 (2007).

\bibitem{Billard}
    J. Billard, E. Figueroa-Feliciano and L. Strigari, \emph{Implication of neutrino backgrounds on the reach of next generation dark matter direct detection experiments}, Phys. Rev. D {\bf 89}, 023524 (2014).

\bibitem{Grant0}
    I. P. Grant, \emph{Relativistic self-consistent fields}, Proc. R. Soc. London Ser. A {\bf 262}, 555-576 (1961).

\bibitem{Desclaux}
    J. P. Desclaux, \emph{A multiconfiguration relativistic Dirac-Fock program}, Comput. Phys. Commun. {\bf 9}, 31-45 (1975). 

\bibitem{Grant}
    K. G. Dyall, I. P. Grant, C. T. Johnson, F. A. Parpia and E. P. Plummer, \emph{GRASP: A general-purpose relativistic atomic structure program}, Comput. Phys. Commun. {\bf 55}, 425-456 (1989).

\bibitem{Visscher}
    L. Visscher and K. G. Dyall, \emph{Dirac¨CFock atomic electronic structure calculations using different nuclear charge distributions}, At. Data and Nucl. Data Tables {\bf 67}, 207-224 (1996).

\bibitem{Gu}
    P. Gu, \emph{Research on low-momentum transfer Compton scatterings}, Dissertation for master degree, Sichuan University: China (2019, in Chinese).

\bibitem{Experiment}
    P. Gu \emph{et al.}, \emph{Siguature on Compton Scattering at Low Momentum Transfer in Germanium for Dark Matter Searches}, (in parparation).

\bibitem{Pratt1}
    P. M. Bergstrom, T. Suri\'c, K. Pisk and R. H. Pratt, \emph{Compton scattering of photons from bound electrons: Full relativistic independent-particle-approximation calculations}, Phys. Rev. A {\bf 48}, 1134-1162 (1993).

\bibitem{Pratt}
    P. M. Bergstrom and R. H. Pratt, \emph{An overview of the theories used in Compton Scattering Calculations}, Radiat. Phys. Chem. {\bf 50}, 3-29 (1997).

\bibitem{Kaplan}
    I. G. Kaplan, B. Barbiellini and A. Bansil, \emph{Compton scattering beyond the impulse approximation}, Phys. Rev. B {\bf 68}, 235104 (2003).

\bibitem{Suric}
    T. Suri\'c, \emph{Compton scattering beyond impulse approximation: Correlation, nonlocal-exchange and dynamic effects}, Radiat. Phys. Chem. {\bf 75}, 1646-1650 (2006).

\bibitem{Pratt2}
    R. H. Pratt, L. A. LaJohn, V. Florescu, T. Suri\'c, B. K. Chatterjee and S. C. Roy, \emph{Compton scattering revisited}, Radiat. Phys. Chem. {\bf 79}, 124-131 (2010).

\bibitem{Drukarev1}
    E. G. Drukarev, A. I. Mikhailov and I. A. Mikhailov, \emph{Low-energy K-shell Compton scattering}, Phys. Rev. A {\bf 82}, 023404 (2010).

\bibitem{Drukarev2}
    E. G. Drukarev and A. I. Mikhailov, \emph{High energy atomic physics}, (Springer International Publishing, Switzerland 2016).

\bibitem{Qiao}
    C.-K. Qiao, H.-C Chi, L. Zhang, P. Gu, C.-P. Liu, C.-J Tang, S.-T. Lin and K.-N Huang, \emph{Relativistic Impulse Approximation in Compton Scattering}, J. Phys. B: At. Mol. Opt. Phys {\bf 53}, 075002 (2020) arXiv:1902.02301[physics.atom-ph].

\end{thebibliography}
\end{document}